# Intrinsic Valley Polarization and Anomalous Valley Hall Effect in Single-Layer 2H-FeCl$_2$

Pei Zhao,[1] Yandong Ma,[1,]* Hao Wang,[1] Baibiao Huang,[1] Liangzhi Kou[2] and Ying Dai[1,]*

[1]School of Physics, State Key Laboratory of Crystal Materials, Shandong University, Shandanan Str. 27, Jinan 250100, People's Republic of China

[2]School of Chemistry, Physics and Mechanical Engineering Faculty, Queensland University of Technology, Garden Point Campus, QLD 4001, Brisbane, Australia

*Corresponding author: yandong.ma@sdu.edu.cn (Y. M); daiy60@sina.com (Y. D)

## Abstract

Valley, as a new degree of freedom for electrons, has drawn considerable attention due to its significant potential for encoding and storing information. Lifting the energy degeneracy to achieve valley polarization is necessary for realizing valleytronic devices. Here, on the basis of first-principles calculations, we show that single-layer FeCl$_2$ exhibits a large spontaneous valley polarization (~101 meV) arising from the broken time-reversal symmetry and spin-orbital coupling, which can be continuously tuned by varying the direction of magnetic crystalline. By employing the perturbation theory, the underlying physical mechanism is unveiled. Moreover, the coupling between valley degree of freedom and ferromagnetic order could generate a spin- and valley-polarized anomalous Hall current in the presence of the in-plane electric field, facilitating its experimental exploration and practical applications.

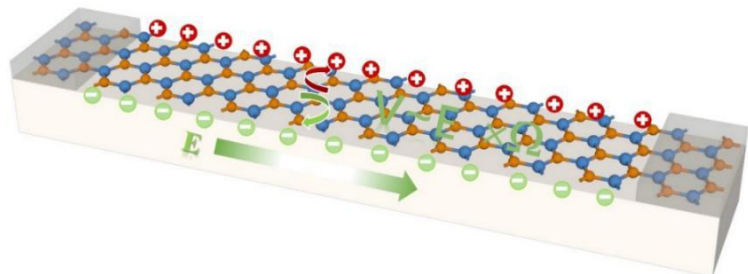

# 1. Introduction

In analogy to charge for electronics and spin for spintronics, utilizing valley degree of freedom to encode and process information is characterized as valleytronics, which is an important hotspot for developing next-generation minimized devices.[1-3] The valley refers to a local energy minimum (maximum) in conduction (valence) band at the K and K′ points, which is robust against phonon and impurity scatterings due to the large separation in the momentum space. The exploration for valleytronic materials has been restricted until the recent emergency of two-dimensional (2D) systems with a hexagonal lattice. The first proposal for realizing 2D valleytronics is reported in graphene,[4, 5] however, the inversion symmetry and weak spin-orbit coupling (SOC) result in degenerated character at its six valleys, limiting its valleytronic applications. Subsequently, single-layer (SL) transitional metal dichalcogenides (TMDs) are identified to harbor the valley physics,[6-10] which arises from the broken inversion symmetry combined with strong SOC originated from the *d* orbitals of the metal atoms. The interesting coupled spin and valley physics make SL TMDs as the most competitive candidates to be applied in valleytronic devices. Besides these two typical systems, many other SL structures are proposed to exhibit the valley physics, such as two-dimensional Tl/SiC and $Tl_2O$.[11-13]

Currently, the major challenge in 2D valleytronics is to break the degeneracy between the valleys, thus producing the valley polarization, which is indispensable for valleytronic applications. Although under circularly polarized light irradiation, the valley polarization is already experimentally achieved in SL $MoS_2$,[1] the complicated tuning equipment and short luminescence lifetime make it a challenging task in practical applications. An alternative way to reach the valley polarization is to break time reversal symmetry,[14-22] namely introducing ferromagnetism. When extra electrons/holes are introduced in such system, the Fermi level can be shifted between the K and K' valleys, thus achieving inequivalent carrier occupation in K and -K valleys. Actually, several approaches have been proposed, including magnetic proximity effect and magnetic doping. However, these external methods suffer from many drawbacks. For example, for the magnetic proximity effect caused by ferromagnetic substrates,[16,17] it usually deforms the host band structures and, in turn, degrades the valleytronic performances. While for magnetic doping, the doped atoms are easy to assemble into clusters, which will scatter the carriers, thus being unfavorable for valleytronics. To solve these problems, spontaneous valley polarization is recently proposed, which have been reported in MnPSe$_3$ and VSe$_2$.[23-25] In these systems, the valley polarization is produced by the intrinsic ferromagnetism, without needing any external tuning. Though highly valuable, the spontaneous valley polarization reported in these systems are rather small, greatly hindering their applications. Except these two systems, no other systems are reported to exhibit the spontaneous valley polarization, largely due to the fact that 2D ferromagnetic semiconductor is rare itself. Therefore, searching for 2D ferromagnetic semiconductors with spontaneous valley polarization is still highly desirable.

Since 1950s, bulk FeCl$_2$ has been synthesized and it has been demonstrated to be antiferromagnets, in which the interlayer metal ions favor antiferromagnetic orderings, while intralayer metals ions favor ferromagnetic orderings.[26] Recently, SL FeCl$_2$ has been identified to be a ferromagnetic half-metal with 1T phase and a ferromagnetic semiconductor with 2H phase. And the formation energy of SL 1T-FeCl$_2$ is about 0.05 eV/atom lower than that of SL 2H-FeCl$_2$.[27-29] Although the SL 1T-FeCl$_2$ is found to be more stable, the stability of 2H-FeCl$_2$ has been confirmed as well, indicating its feasible in future experiments.[30, 31]

Here, we propose a novel 2D valleytronic material with spontaneous valley polarization in SL FeCl$_2$ based on first-principles calculations. It is found that SL FeCl$_2$ is a ferromagnetic semiconductor with a direct band gap locating at the K valleys. The Fe$^{2+}$ ions in SL FeCl$_2$ favors a high-spin configuration, which leads to a large magnetic moment of 4 $\mu_B$ on each Fe atom. The magnetocrystalline anisotropy energy (MAE) is estimated to be 0.21 meV per atom, indicating the magnetization orientation is perpendicular to the 2D plane. More importantly, without any external tuning, a large spontaneous valley polarization of ~101 meV at the K and K′ valleys is obtained in SL FeCl$_2$, arising from the coexistence of strong SOC and exchange interaction of localized $d$ electrons of Fe atoms. The Berry curvature with opposite signs at the K and K′ valleys will lead to the observation of anomalous valley Hall effect in SL FeCl$_2$. Additionally, the valley polarization can be continuously tuned by changing the magnetized direction. When the magnetized direction is reversed from upward to downward, the valley at K and K′ points will be flipped.

## 2. Methods

All calculations are performed based on density functional theory implemented in the Vienna ab initio simulation package (VASP).[32] The cutoff energy is set to 500 eV for plane-wave expansion. The Brillouin zone (BZ) is meshed by a 13×13×1 Monkhorst-Pack grid. To avoid the interaction between adjacent layers, a vacuum space of about 18 Å normal to 2D plane is applied. The exchange and correlation function is treated by the generalized gradient approximation (GGA) in form of Perdew-Burke-Ernzerhof (PBE) functional.[33] The projected argument wave (PAW) potential was used to describe the ion–electron potential.[34] The HSE06 functional is employed to obtain the accurate band structures.[35] Spin polarization and SOC are included by noncollinear calculations. The convergence accuracy of total energy and Hellmann-Feynman force are respectively set as $10^{-5}$ eV and 0.01 eV/Å. The MAE is defined as MAE = $(E_{100}-E_{001})/n$, of which $E_{100}$ and $E_{001}$ represent the total energies with magnetocrystalline directions along in-plane and out-of-plane directions, respectively, and $n$ is the number of atoms. The phonon spectrum is obtained by using a 5×5×1 supercell based on density functional perturbation theory (DFPT) method.[36] For Berry curvature of SL $FeCl_2$, the maximally localized Wannier function method implemented in WANNIER90 package is employed,[37] and p orbitals of Cl and d orbitals of Fe are chosen as projected orbitals.

## 3. Results and Discussion

The crystal structure of SL H-$FeCl_2$ is presented in **Figure 1(a)**. We can see that one Fe atomic layer is sandwiched between two Cl atomic layers and each Fe atom is coordinated by six Cl atoms, forming a hexagonal lattice with the space group $D_{3h}$. Therefore, the inversion symmetry of SL $FeCl_2$ is broken, which is essential to realize the valley polarization. The calculated lattice constant is 3.36 Å and the bond length of Cl-Fe is 2.47 Å. To characterize the bonding character in SL $FeCl_2$, we plot its electron localization function (ELF) in **Figure 1(b)**. Clearly, the electrons are mainly localized around the Fe and Cl atoms, suggesting a typical ionic bonding between Fe and Cl. We then examine its stability by performing phonon calculations. As shown in **Figure 1(c)**, the absence of imaginary frequency in the whole BZ demonstrates its dynamic stability.

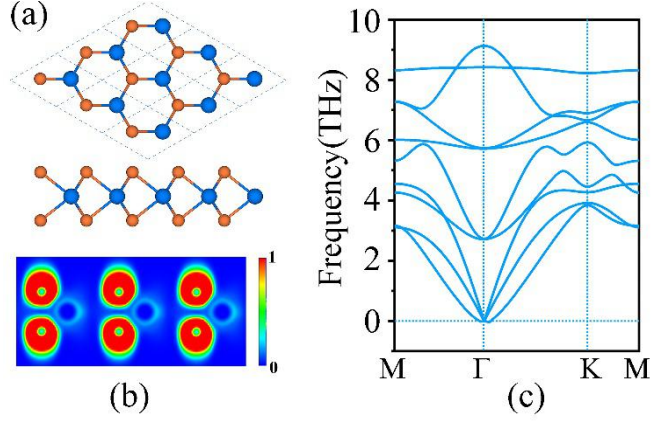

**Figure 1**. (a) Top and side views of the crystal structure of SL 2H-FeCl$_2$. (b) Electron localization function (ELF) of SL 2H-FeCl$_2$. (c) Phonon spectrum of SL 2H-FeCl$_2$.

Our electronic calculations show that SL FeCl$_2$ is spin polarized, and each unit cell holds a total magnetic moment of 4 μ$_B$. Moreover, the orbit moment on each Fe atom is found to be 0.01 μ$_B$. To explore the origin of the magnetic moment, we plot the spin polarized charge density of SL FeCl$_2$ in **Figure 2(a)**. We can see that the magnetic moment is mainly carried by Fe atoms. It should be noted that the existence of magnetic moments does not guarantee the formation of magnetic coupling between them. To that end, we construct a 2×2 supercell with four Fe atoms to investigate its magnetic ground state. We consider two magnetic configurations, namely, FM [**Figure 2(b)**] and AFM [**Figure 2(c)**] states. The magnetic ground state of SL FeCl$_2$ is found to be FM, with an energy difference of 50 meV lower than AFM state. Therefore, SL FeCl$_2$ is a ferromagnetic material. Besides, We employ the mean-field theory[38, 39] $K_B T_c = (2/3)\Delta E$ to estimate the Curie temperature ($T_c$) of SL 2H-FeCl$_2$, $K_B$ is the Boltzmann constant, $\Delta E$ is the quarter of energy difference between ferromagnetic and antiferromagnetic configurations in a 2×2 supercell. The $T_c$ for SL 2H-FeCl$_2$ is found to be about 96 K.

To deeply understand the magnetic properties of SL FeCl$_2$, we analyze the splitting of $d$ orbitals of Fe atom. As mentioned above, each Fe atom is coordinated with six Cl atoms, forming a trigonal prismatic structure. Under such a crystal field,[7] the five $d$ orbitals of Fe atom split into three groups: $A_1$ ($d_{z^2}$), $E_1$ ($d_{xy}$, $d_{x^2-y^2}$) and $E_2$ ($d_{xz}$, $d_{yz}$). Due to that the $d$ electrons in Fe$^{2+}$-based systems generally favor the high-spin configuration, the unpaired electrons with the same spin direction are assigned to maximum numbers. Considering that the electron configuration of $Fe^{2+}$ is $d^6$, the $A_1$ orbital is fully occupied by two electrons with opposite spins, $E_1$ ($E_2$) orbitals are half-filled by two spin-up electrons. When further including the influence of magnetic exchange field, the orbitals further split into spin-up and spin-down groups, as shown in **Figure 2(d)**. This leads to a magnetic moment of 4 $\mu_B$ on each Fe atom. These results are in consistent with the DOS shown in **Figure 2(e)**. Moreover, we find that the optimized Fe-Cl-Fe angle is 87°, which is near the ideal 90°. According to the Goodenough-Kanamori rules,[40-42] 90° bond angle usually associated with FM ordering. With this result in hand, we can understand why FM coupling is favorable in SL $FeCl_2$.

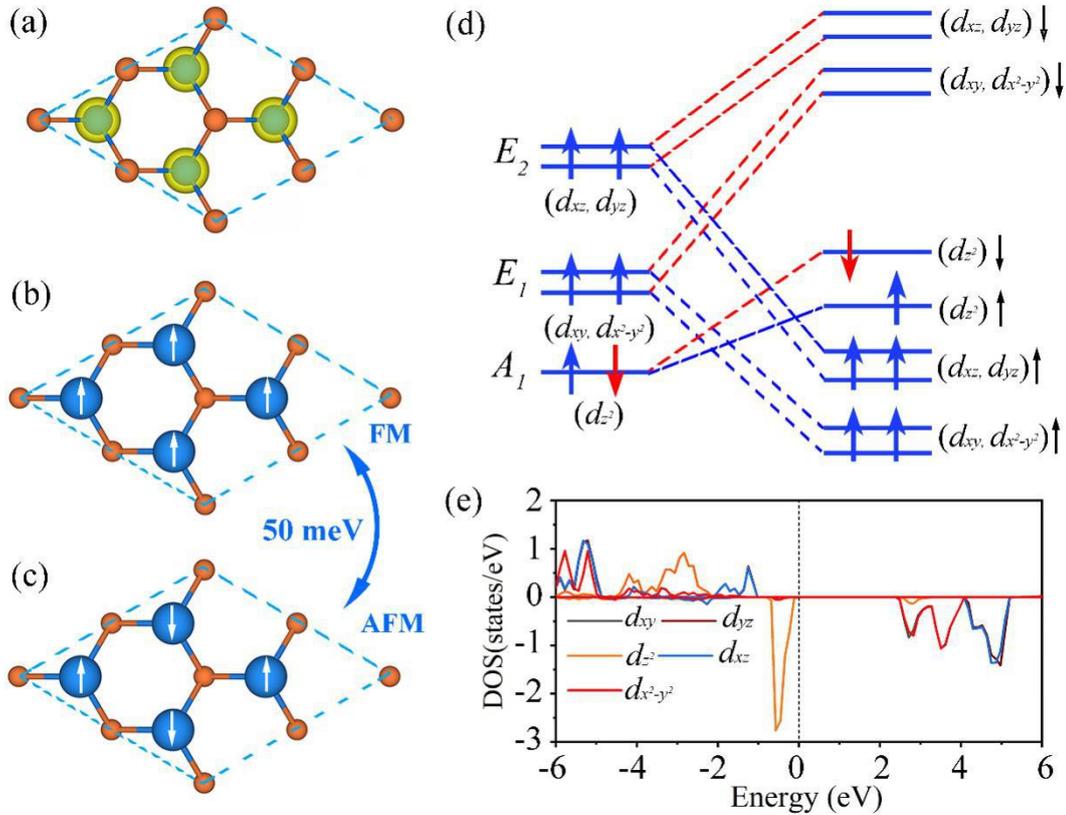

**Figure 2**. (a) Spin polarized charge density of SL $FeCl_2$ (b) FM and (c) AFM configurations of SL $FeCl_2$. (d) The splitting of d orbitals of Fe atoms under trigonal prismatic crystal field. (e) Projected density of states of Fe-d orbitals of SL $FeCl_2$.

Having estimated the magnetic behaviors of SL $FeCl_2$, we then explore its electronic properties. The band structure and density of states of SL $FeCl_2$ without considering SOC is shown in **Figure 3(a)**, from which we can see that it is a ferromagnetic semiconductor. Another interesting point we can see from **Figure 3(a)** is that the valance band maximum (VBM) and conduction band minimum (CBM) of SL $FeCl_2$ locate at the K and K′ points, forming two valleys. SL $FeCl_2$ thus is a promising 2D valleytronic semiconductor. Upon switching on SOC, interestingly, the energetic degeneracy between the K and K′ valleys are lifted, giving rise to a valley polarization in SL $FeCl_2$. It is worth noting that the valley polarization is different from most of previous proposals where the polarization is induced by external approaches, i.e., dopants, magnetic substrates and chirality-dependent lights. Here, the valley polarization occurs spontaneously, without needing any external modulation. We define intensity of valley polarization as $\Delta E = E_K - E_{K'}$. Remarkably, the spontaneous valley polarization in the conduction bands of SL $FeCl_2$ is as large as 101 meV (**Figure S1** and **Figure 3(b)**), which can resist the annihilation of valley polarization caused by kinetic energy at room temperature, being essential to practical applications in valleytronics. Different from the case of conduction band, the valley polarization in the valence bands is rather small, which is only 4 meV. It is interesting to note that the occurrence of valley polarization in SL $FeCl_2$ does not guarantee the valley polarization in its bulk form, which sensitively depends on the stacking pattern and magnetic coupling type.

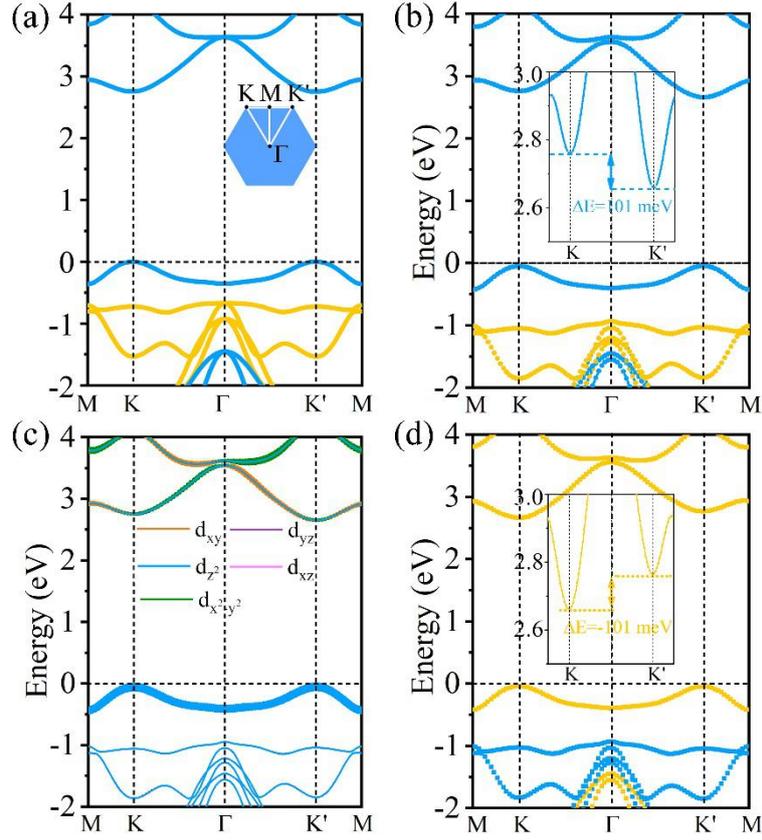

**Figure 3**. (a) Spin polarized band structure of SL FeCl$_2$ without SOC based on HSE06 functional. Insert in (a) shows the 2D Brillouin zone with marking the high-symmetry points. (b) Band structure of SL FeCl$_2$ with SOC with magnetic moment along +z direction. (c) Orbital-resolved band structures of SL FeCl$_2$ with SOC. (d) is same as (b) but with magnetic moment along -z direction. The orange and blue lines represent spin-up and spin-down states, respectively. The Fermi level is shifted to the VBM.

We then explore the discrepancy between the valley polarizations of the conduction and valance bands. As shown in **Figure S2**, the CBM and VBM of SL FeCl$_2$ are mainly contributed by Fe atoms. In detail, the CBM of SL FeCl$_2$ is mainly from $d_{xy}$ and $d_{x^2-y^2}$ orbitals, and VBM is mainly dominated by $d_{z^2}$ orbital and a fraction of $d_{xy}$ and $d_{x^2-y^2}$ orbitals; see **Figure 2(e)** and **3(c)**. As SOC leads to the valley polarization, it can be approximated to be contributed by the intra-atomic interaction $\hat{L} \bullet \hat{S}$, the SOC term can be written as $\hat{H}_{SOC}^0 + \hat{H}_{SOC}^1 = \lambda \hat{L} \bullet \hat{S}$ .[43, 44] Here, $\hat{L}$ and $\hat{S}$ represent the orbital angular moment and spin angular moment, respectively. $\hat{H}_{SOC}^0$ is the interaction between the same spin states and $\hat{H}_{SOC}^1$ indicates the interaction

between opposite ones. $\hat{H}_{SOC}^0$ and $\hat{H}_{SOC}^1$ can be expressed in following forms:

$$\hat{H}_{SOC}^0 = \lambda \hat{S}_z (\hat{L}_z \cos\theta + \frac{1}{2}\hat{L}_+ e^{-i\phi}\sin\theta + \frac{1}{2}\hat{L}_- e^{+i\phi}\sin\theta),$$

$$\hat{H}_{SOC}^1 = \frac{\lambda}{2}(\hat{S}_+ + \hat{S}_-)(-\hat{L}_z \sin\theta + \frac{1}{2}\hat{L}_+ e^{-i\phi}\cos\theta + \frac{1}{2}\hat{L}_- e^{+i\phi}\cos\theta),$$

Because the spin degeneracy between the spin-up and spin-down bands is significantly broken in SL FeCl$_2$, there is almost no the interaction between spin-up and spin-down states near the Fermi level. Thus, the contribution from $\hat{H}_{SOC}^1$ can be ignored. As magneto-crystalline direction in SL FeCl$_2$ is perpendicular to the 2D plane ($\theta=0°$), $\hat{H}_{SOC}^0$ term can be simplified as $\hat{H}_{SOC}^0 \propto \alpha \cos\theta \hat{L}_z$. Accordingly, the Hamiltonian for SOC is determined by the orbital angular momentum operator $\hat{L}_z$. The group symmetry of the wave vector at the valleys is $C_{3h}$, and the symmetry adapted basis functions are chosen as

$$|\phi_c^\tau\rangle = \sqrt{\frac{1}{2}}(|d_{x^2-y^2}\rangle + i\tau |d_{xy}\rangle) \text{ and } |\phi_v\rangle = |d_{z^2}\rangle,$$

where the subscript $c$ and $v$ represent conduction and valence bands, respectively. The superscript $\tau$ indicates the valley index ($\tau = \pm 1$). The SOC effect on conduction band and valence band can be expressed by

$$\langle\phi_c^\tau|\hat{H}_{soc}^0|\phi_c^\tau\rangle = \sqrt{\frac{1}{2}}(\langle d_{x^2-y^2}|\hat{H}_{soc}^0|d_{x^2-y^2}\rangle \mp \langle d_{xy}|\hat{H}_{soc}^0|d_{xy}\rangle)$$

$$\langle\phi_v|\hat{H}_{soc}^0|\phi_v\rangle = \langle d_{z^2}|\hat{H}_{soc}^0|d_{z^2}\rangle,$$

The energy level of CBM at K and K′ valleys are respectively

$$E_c^k = \sqrt{\frac{1}{2}}\langle d_{x^2-y^2}|\hat{H}_{soc}^0|d_{x^2-y^2}\rangle - \langle d_{xy}|\hat{H}_{soc}^0|d_{xy}\rangle,$$

$$E_c^{k'} = \sqrt{\frac{1}{2}}\langle d_{x^2-y^2}|\hat{H}_{soc}^0|d_{x^2-y^2}\rangle + \langle d_{xy}|\hat{H}_{soc}^0|d_{xy}\rangle,$$

While the energy level of VBM at K and K′ valleys are

$$E_v^k = E_v^{k'} = \langle d_{z^2}|\hat{H}_{soc}^0|d_{z^2}\rangle.$$

As the magnetic quantum number $m_z$ for $|d_{x^2-y^2}, d_{xy}\rangle$ and $|d_{z^2}\rangle$ are 2 and 0, respectively, the energy difference of CBM between K and K′ is $\Delta E_1 = E_c^k - E_c^{k'} \neq 0$, and the energy difference of VBM between K and K′ is $\Delta E_2 = E_v^k - E_v^{k'} = 0$. Accordingly, we can understand why significantly valley polarization occurs at CBM, while the degeneracy between the K and K′ valleys of VBM is almost preserved, as for the values of 4 meV. It should be noted that in this simple model for VBM, we only consider the $d_{z^2}$ orbital for approximation. In fact, VBM is also contributed slightly from $d_{xy}$ and $d_{x^2-y^2}$ orbitals, which results in that the K and K' valleys in VBM would not be strictly degenerate.

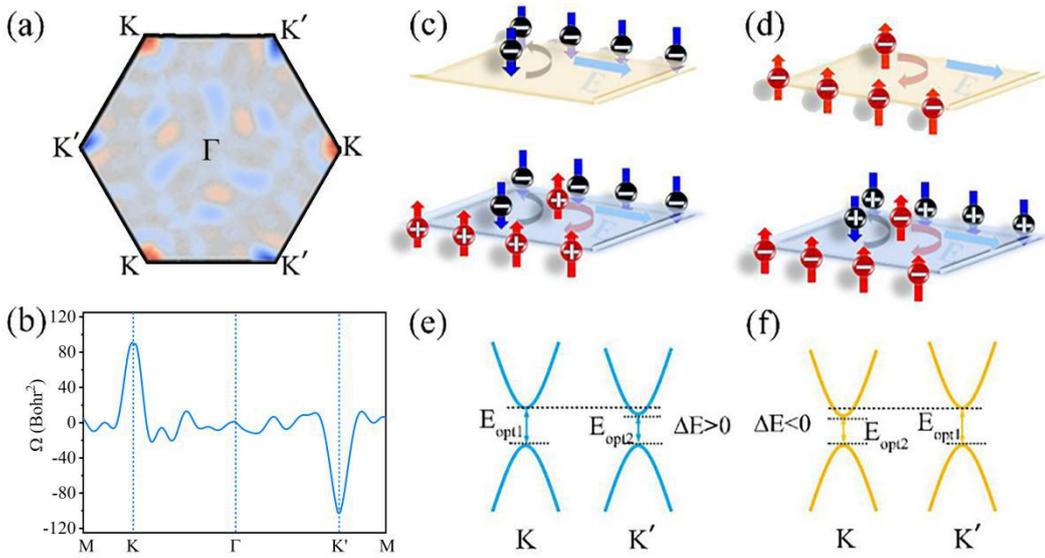

**Figure 4**. (a) Contour map of Berry curvature of SL FeCl$_2$ in the whole 2D BZ (b) Berry curvature of SL FeCl$_2$ along the high-symmetry points. (c) Schematics of anomalous valley Hall effect under electron doping (top) and light-irradiation (bottom) in SL FeCl$_2$. (d) is the same as (c) but with opposite magnetic moment. The blue and red arrows in (c, d) represent spin-down and spin-up states, respectively. (e) Schematic of valley polarization in SL FeCl$_2$. (f) is the same as (e) but with opposite magnetic moment.

After identifying the spontaneous valley polarization in SL FeCl$_2$, it is also important to reveal its Berry curvature which is associated with anomalous velocity of Bloch electrons $\vec{v}_a \sim \vec{E} \times \Omega(\vec{k})$. Here, $\vec{E}$ is external electric field and $\Omega(\vec{k})$ is the Berry curvatures. On the basis of the Kubo formula derivation, the Berry curvature can be written as $\Omega(k) = -\sum_{n}\sum_{n \neq n'} f_n \frac{2\,\text{Im}\langle \psi_{nk}|v_x|\psi_{n'k}\rangle\langle \psi_{n'k}|v_y|\psi_{nk}\rangle}{(E_n - E_{n'})^2}$.[45] Here, the summation is over all the occupied states. $f_n$ is Fermi-Dirac distribution function, $\psi_{nk}$ is Bloch function and $v_x$ ($v_y$) is the velocity operate. Obviously, a sizable $\Omega(\vec{k})$ at the valleys can accelerate transvers transport of carriers, which is key to detect valley Hall current by electric measurements. **Figure 4(a)** and **(b)** display the calculated Berry curvatures of SL $FeCl_2$ as a contour map in the whole 2D BZ and as a curve along high-symmetry points, respectively. Clearly, $\Omega(\vec{k})$ is characterized with unequal values and opposite signs for the K and K′ valleys, while the value is almost zero around the Γ point. Noticeably, $\Omega(\vec{k})$ of SL $FeCl_2$ can reach up to 91 (103) Bohr$^2$ for K (K′) valley, which is larger than most of the previously reported values. When SL $FeCl_2$ is reversed with opposite magnetic moment, the valley polarization will also be reversed [**Figure 3(d)**] and then the corresponding values of the Berry curvatures at the K and K′ valleys will be exchanged. However, the sign of the Berry curvature remains to be the same.

Under such condition, the valley Hall effect can be expected in SL $FeCl_2$. As shown in **Figure 4(c)**, when shifting the Fermi level between the K and K′ valleys, the spin-down electrons at K′ valley will accumulate on the right side in the presence of an in-plane electric field. And when SL $FeCl_2$ is magnetized with opposite magnetic moment, the spin-up electrons at K valley will gain opposite transverse velocities, shifting towards left side of the sample, which is attributed to its opposite Berry curvature; see **Figure 4(d)**. In either of these two scenarios, the electric potential difference between the two sides of the sample is generated, which can be detected as a Hall voltage. To facilitate the experimental exploration, we calculate the anomalous Hall conductivity $\sigma_{xy}$ of SL $FeCl_2$ by integrating the Berry curvature over the 2D Brillouin zone. As shown in **Figure S3**, when the Fermi level is shifted to between the CBM of the K and K′ valleys, as marked by the dashed lines in **Figure S3**, a fully spin- and valley-polarized Hall conductivity will be generated. Meanwhile, the valley selectivity can be manipulated in SL $FeCl_2$ under the light irradiation. **Figure 4(e)** and **(f)** show the schematics of the low-energy band structures of SL $FeCl_2$. When SL $FeCl_2$ is magnetized upward, the energy gap at K′ valley $\Delta_{opt2}$ is 2.70 eV, which is less than that of $\Delta_{opt1}$. Therefore, when the frequency of the incident light ω meet the requirement of $\frac{\Delta_{opt2}}{\hbar} \leq \omega < \frac{\Delta_{opt1}}{\hbar}$, the spin-down electrons and spin-up holes at K′ valley will be generated. With the help of the in-plane electric field, the photogenerated electrons and holes will move towards opposite direction, resulting from the opposite Berry curvatures in the valance band and conduction band; see the bottom panel in **Figure 4(c)**. And similar phenomenon is expected when SL FeCl$_2$ is magnetized opposite [**Figure 4(d)**]. These thus lead to measurable Hall current identified with valley, spin and charges, which is attractive for applications in spintronic devices.

Considering the fact that electronic properties of 2D materials are susceptible to strain,[46-48] one may wonder if the external strain could influence the valley polarization of SL FeCl$_2$. We thus exert -4%-4% biaxial strain to SL FeCl$_2$ to address this concern. Here, we define the strain as $\varepsilon = \frac{a - a_0}{a}$, where $a$ and $a_0$ are the lattice constants of the strained and unstrained SL FeCl$_2$, respectively. From **Figure S4**, we can clearly see that the significant valley polarization is preserved, with the values of ΔE being in the range of 100 meV and 105 meV. Therefore, the valley polarization in SL FeCl$_2$ is robust against the strain, which is favorable for practical valleytronic applications.

Another interesting point we wish to address is the MAE of SL FeCl$_2$. As we mentioned above, SL FeCl$_2$ exhibit a weak MAE of ~0.21 meV per atom. This value, however, is comparable with that of SL CrI$_3$. As the suppression of spin fluctuations in SL CrI$_3$ has been demonstrated in experiment,[49, 50] we believe that the spin fluctuation in SL FeCl$_2$ can also be suppressed. As shown in **Figure S5** and **S6**, the valley polarization is sensitive to the magnetocrystalline direction in SL FeCl$_2$, which renders the conversion of valley polarization between K and K′ valleys possible. Interestingly, when the magnetocrystalline direction is along the in-plane (θ=90°), the K and K′ valleys are degenerated in energy, and thus the valley polarization vanishes. This can also be understood by employing the Hamiltonian for SOC. For $\hat{H}_{SOC}^0 = \lambda \hat{S}_z (\hat{L}_z \cos\theta + \frac{1}{2}\hat{L}_+ e^{-i\phi} \sin\theta + \frac{1}{2}\hat{L}_- e^{+i\phi} \sin\theta)$, when θ=90°, it is simplified as $\hat{H}^0_{SOC} \propto \lambda \hat{L}_{\pm}$. The Hamiltonian for SOC thus is only related to $\hat{L}_{\pm}$. The SOC effect on conduction band can be expressed by, $\langle \phi^\tau_c | \hat{H}^0_{soc} | \phi^\tau_c \rangle = \sqrt{\frac{1}{2}}(\langle d_{x^2-y^2} | \hat{H}^0_{soc} | d_{x^2-y^2} \rangle \mp \langle d_{xy} | \hat{H}^0_{soc} | d_{xy} \rangle)$. Due to the ladder operator,

$$\hat{L}_- | d_{x^2-y^2} \rangle = \hat{L}_- | d_{xy} \rangle \propto | d_{xz}, d_{yz} \rangle,$$

$$\langle \phi^\tau_c | \hat{H}^0_{soc} | \phi^\tau_c \rangle = \gamma (\langle d_{x^2-y^2} | d_{xz}, d_{yz} \rangle \mp \langle d_{xy} | d_{xz}, d_{yz} \rangle) = 0.$$

Thus, the energy level difference between K and K′ valleys is $\Delta E = E^k_c - E^{k'}_c = 0$.

## 4. Conclusion

In summary, we propose a novel 2D ferromagnetic semiconductor, SL FeCl$_2$, with a direct band gap of 2.7 eV. The Fe$^{2+}$ ions in SL FeCl$_2$ favors a high-spin configuration, resulting in a large spin-polarization with a magnetic moment of 4 μ$_B$ per unit cell. Moreover, SL FeCl$_2$ is identified to be a 2D valleytronic material. Different form previously reported 2D valley materials, SL FeCl$_2$ holds a spontaneous valley polarization which can reach up to 100 meV in the conduction band. Such a large valley polarization is sizeable enough to resist the annihilation of valley polarization caused by kinetic energy at room temperature. The underlying mechanism for the discrepancy between the valley polarizations of the conduction and valance bands is also revealed. In addition, the valley polarization in SL FeCl$_2$ is robust against external strain, which is favorable for its applications in practical valleytronic devices. It should be noted that these findings can be generalized to other 2D ferromagnetic materials with the following three ingredients: (1) its inversion symmetry should be broken; (2) its time reversal symmetry should be broken; (3) its CBM (VBM) should be located at the K points. Our results not only reveal the novel valley physics in SL FeCl$_2$ but also provide a platform for practical spintronic and valleytronic applications.

*Note added.* We noticed one similar work on the prediction of 2H-FeCl$_2$ monolayer as ferrovalley materials which was submitted to arXiv at almost the same time as ours[51].

## Acknowledgement

This work is supported by the National Natural Science Foundation of China (No.